\documentclass[onecolumn,showpacs,preprintnumbers,amsmath,amssymb,ajp]{revtex4-1}

\usepackage{epsfig}
\usepackage{graphicx}
\usepackage{dcolumn}
\usepackage{bm} 

\newcommand{\beq}{\begin{equation}}
\newcommand{\eeq}{\end{equation}}
\newcommand{\beqa}{\begin{eqnarray}}
\newcommand{\eeqa}{\end{eqnarray}}

\newcommand{\ket}[1]{|#1\rangle}                
\newcommand{\bra}[1]{\langle#1|}                

\def\ajp#1{{ Am.\ J.\ Phys.} {\bf #1}}

\def\jmo#1{{ J.\ Mod.\ Opt.} {\bf#1}}

\def\jpb#1{{ J.\ Phys.\ B} {\bf#1}}
\def\pra#1{{ Phys.\ Rev. A\/} {\bf#1}}

\def\prl#1{{ Phys.\ Rev.\ Lett.} {\bf#1}}

\begin{document}


\title{Maximizing genuine multipartite entanglement of $N$ mixed qubits}

\author{S. Agarwal and S.M. Hashemi Rafsanjani}

\affiliation{ Rochester Theory Center and the Department of Physics
\& Astronomy\\
University of Rochester, Rochester, New York 14627}



\date{\today}

\begin{abstract}
Beyond the simplest case of bipartite qubits, the composite Hilbert space of multipartite systems is largely unexplored. In order to explore such systems, it is important to derive analytic expressions for parameters which characterize the system's state space. Two such parameters are the degree of genuine multipartite entanglement and the degree of mixedness of the system's state. We explore these two parameters for an $N$-qubit system whose density matrix has an $X$ form. We derive the class of states that has the maximum amount of genuine multipartite entanglement for a given amount of mixedness. We compare our results with the existing results for the $N=2$ case. The critical amount of mixedness above which no $N$-qubit $X$-state possesses genuine multipartite entanglement is derived. It is found that as $N$ increases, states with higher mixedness can still be entangled.
\end{abstract}



\maketitle

\section{Introduction}

Quantification of entanglement of a bipartite state with arbitrary dimensions is a challenging problem and solutions are known only for some special cases with systems consisting of two qubits \cite{Wootters-98}, a qubit and a qutrit \cite{Peres-96, Horodecki-96}, two qudits in a highly symmetric state \cite{Horodecki-99,Vollbrecht-01}, a double gaussian state \cite{Simon-00}, a pure bipartite state \cite{Ekert-95,Grobe-etal-94}, etc. 
Beyond the bipartite case, quantifying the entanglement of a system consisting of three or more parties becomes even more challenging. For such systems, it becomes important to distinguish between genuinely multipartite entangled states, for which all the parties are entangled with every other party, from the bi-separable states for which some parties might be entangled with each other while being separable from the rest \cite{Guhne-10, Ma-11, Jungnitsch-11}. 

In \cite{Ma-11} it was shown that a generalization of Wootters concurrence, which was derived for the two-qubit case \cite{Wootters-98}, is a good measure to quantify genuine multipartite entanglement. Although various witness operations to verify genuine multipartite entanglement existed \cite{Toth-05, Villar-06, Toth-09, Gao-10, Bancal-11, Huber-11}, it was not until recently that an analytic expression for $C_{GM}$ was derived \cite{Hashemi-12}. It was shown in \cite{Hashemi-12} that if a state of an $N$-qubit system has an $X$ form \cite{Yu-07}, one can get  $C_{GM}$ analytically. 

The $X$-states are density matrices whose only non-zero elements are the diagonal or the anti-diagonal elements when written in an orthonormal product basis of the $N$-qubit system. Although the properties of $X$ density matrices for two-qubit systems are extensively studied in the literature \cite{Yu-07, Quesada-12, Hashemi-12Ap}, the properties of $N$-qubit $X$ density matrices are still largely unexplored. In this report, we investigate the properties of these $X$-states of $N$-qubit systems. In general, an $X$-state of $N$-qubits has $(2^{N+1}-1)$ independent real parameters. Instead of addressing this exponentially growing parameter space, we restrict our attention only to two fundamental quantities, genuine multipartite concurrence and linear entropy. Linear entropy is a measure of mixedness and for a given state, say $\rho$, linear entropy is defined as 
\begin{align}\label{e.S}
S(\rho)=\frac{d}{d-1}(1-\mathrm{Tr}(\rho^2)),
\end{align}
where $d$ is the dimensionality of $\rho$. The linear entropy ranges from $0$ to $1$ with $0$ indicating $\rho$ to be a pure state and $1$ indicating $\rho$ to be completely mixed. In \cite{Munro-01, Wei-03}, Munro \textit{et al.} explored the concurrence-entropy relations for all physically allowed two-qubit states. They found the class of states that has the maximum amount of entanglement for a given degree of entropy. It was shown that as the entropy increases, the maximum achievable amount of entanglement by any possible bipartite qubit state decreases. It was also shown in \cite{Munro-01, Wei-03, Zyczkowski-98} that beyond a critical value of $S_{cr}=8/9$, no two-qubit state could possibly be entangled. The class of states that have the maximum amount of entanglement for a given degree of entropy is referred to as \textit{maximally entangled mixed states} (MEMS) \cite{Munro-01}. For the two-qubit case, the MEMS class happens to be of the $X$ form. This fact suggests that even for an $N$-qubit system, they may be of the $X$ form.

Here, restricting our analysis to the class of $X$-states, we will generalize the two-qubit results to a general $N$-qubit setup. In particular, we find the class of $X$ density matrices that, for a given amount of entropy, has the maximum amount of genuine $N$-qubit entanglement (see Eq. (\ref{e.MEMXS})). We find that the critical amount of entropy, as a function of $N$, above which no $X$ density matrix can possibly exhibit genuine multipartite entanglement is
\begin{align}
S_{cr}(N)=\frac{2^{2N-1}}{(2^N-1)(2^{N-1}+1)}.
\end{align}
We see that this critical entropy increases as a function of $N$. This indicates that in the limit of a large number of qubits, it is possible to find highly mixed states that can still possess entanglement. The dependence of various physical parameters on the entropy of a bipartite system state and the scaling of these dependences on the dimensionality of the two parties involved can be found in \cite{Zyczkowski-98, Bose-00, Adesso-04, Qasimi-11}.

In Section \ref{s.X-Matrices}, we briefly explain the concept of genuine multipartite entanglement. In this section, we also describe the $X$ density matrices of $N$ qubits and the genuine multipartite concurrence that quantifies the entanglement for these states. In Section \ref{s.MEMXS}, we derive the maximally entangled mixed $X$ states (X-MEMS) for $N$ qubits. We conclude in Section \ref{s.Conclusion}.   

\section{Genuine multipartite concurrence of $N$-qubit $X$ density matrices}\label{s.X-Matrices}
A system consisting of $N$ qubits is said to possess genuine multipartite entanglement if each qubit is entangled with every other qubit and not only to some of them \cite{Toth-09}. To make this statement more explicit, we first define what is meant by a bi-separable state: A pure $N$-qubit state, say $\ket{\psi}$, is said to be bi-separable if it can be written as a product of two multi-qubit states: $\ket{\psi_{bs}}=\ket{\psi_1}\otimes\ket{\psi_2}$ (the subscript stands for bi-separable). A mixed state is bi-separable if it can be written as a mixture of bi-separable pure states: $\rho_{bs}=\sum_ip_{i}\ket{\psi^{i}_{bs}}\bra{\psi^{i}_{bs}}$. If an $N$-qubit state is not bi-separable, it is genuinely $N$-partite entangled \cite{Guhne-10, Ma-11, Jungnitsch-11}. 

One of the proposed measures to quantify genuine multi-partite entanglement is genuine multi-partite concurrence \cite{Ma-11}. For a pure state, $\ket{\psi}$, genuine multi-partite concurrence is defined as:
\begin{align}
C_{GM}(\ket{\psi}):=\min_{\gamma\in\Gamma}\sqrt{2(1-\mathrm{Tr}(\rho_{A_{\gamma}}^2))},
\end{align}
where $\Gamma$ represents the set of all possible bi-partitions, $\left\{A_\gamma|B_\gamma\right\}$, of the $N$ parties and $\rho_{A_{\gamma}}$ is the reduced density matrix: $\rho_{A_{\gamma}}=\mathrm{Tr}_{B_{\gamma}}\left(\ket{\psi}\bra{\psi}\right)$. For a mixed state, $\rho$, the concept of $C_{GM}$ can be formally generalized to
\begin{align}
C_{GM}(\rho)=\inf_{\left\{p_i,\ket{\psi^i}\right\}}\sum_ip_i\,C_{GM}(\ket{\psi^i}),
\end{align}
where the minimization has to be carried out over all possible pure state decompositions of the density matrix: $\rho=\sum_ip_i\ket{\psi^i}\bra{\psi^i}$. This minimization procedure renders analytic or even numerical parameterization of $C_{GM}$ infeasible in general. 

For some cases, lower bounds on $C_{GM}$ are available for analysis \cite{Toth-05, Villar-06, Toth-09, Gao-10, Bancal-11, Huber-11}. Only recently, by generalizing the result for the two-qubit case \cite{Wootters-98}, an exact analytic expression for $C_{GM}$ was derived for an $N$-qubit system whose density matrix has an $X$ form \cite{Hashemi-12}. To briefly elaborate on the result in \cite{Hashemi-12}, let us look at an $N$-qubit $X$ density matrix. Arranging the rows and columns in an orthonormal product basis, $\ket{1,1,\cdots,1,1}$, $\ket{1,1,\cdots,1,0}$, $\cdots$, $\ket{0,0,\cdots,0,0}$, the $X$ density matrices have the general structure of the following form:
\begin{align}\label{e.X}
X=
\begin{pmatrix}
a_1&&&&&&&z_1\\
&a_2&&&&&z_2&\\
&&\ddots&&&\reflectbox{$\ddots$}&&\\
&&&a_n&z_n&&&\\
&&&z_n^*&b_n&&&\\
&&\reflectbox{$\ddots$}&&&\ddots&&\\
&z_2^*&&&&&b_2&\\
z_1^*&&&&&&&b_1\\
\end{pmatrix},
\end{align}
where $n=2^{N-1}$. For the $X$ matrix to be a valid density matrix, one must have $\sum_i (a_i+b_i)=1$ and $|z_i|\leq\sqrt{a_ib_i}$. These conditions are necessary and sufficient for the $X$ matrix to have unit trace and positive eigenvalues. In \cite{Hashemi-12}, it was shown that the $C_{GM}$ for the above $N$-qubit state is given by
\begin{align}\label{e.C_gm}
C_{GM}(X)=2\,\max\left\{0,|z_i|-\sum_{j\neq i}^n\sqrt{a_jb_j}\right\}, \quad i=1,2,\cdots,n.
\end{align} 
Note that for the two-qubit case, this result reduces to the previously known result derived by Wootters \cite{Wootters-98}. 

If any one of the qubit's state is traced out from the $X$-matrix, the rest of the qubits become separable. This fact can be understood by noticing that if one traces any of the qubits from Eq. (\ref{e.X}), one gets a diagonal and hence separable reduced density matrix. An example of $X$-states is the class of well known GHZ states. The GHZ states have theoretical and practical importance. In \cite{Mermin-90}, it was shown that there are no bounds to the amount by which the GHZ states can violate the limits imposed by a Bell's inequality. In the context of trapped ions, it has been possible to experimentally create GHZ states of up to fourteen qubits \cite{Monz-11}. With the advances in experimental circuit QED, it is predicted that large number of qubits can be initialized in the GHZ states \cite{Bishop-09}. For two-qubit systems, the $X$-states are central to understanding the relations between fully entangled fractions and entanglement \cite{Grondalski-02}. 

Several important remarks may be made about the experimentally common action of uncontrolled environmental influences on the system. The possibility that an arbitrary initial pure state de-cohering into an $X$-state was studied in \cite{Quesada-12}. Further we can rely on the fact that the $X$-state form itself is robust in the sense that states initially in the $X$ form remain in the $X$ form if local damping channels act on the various non-interacting qubits \cite{Yu-07, Hashemi-12}. Another key point is that decoherence processes that add additional non-$X$ elements cannot decrease the degree of entanglement. That is, the $X$ part alone of a general density matrix provides a lower bound of genuine multipartite entanglement \cite{Ma-11,Hashemi-12Ap}. Thus, we see that $X$-states are an important and experimentally viable class of physically allowed $N$-qubit states.

\section{Maximally entangled mixed $X$-states for $N$-qubits}\label{s.MEMXS}
\begin{figure}[h]
\includegraphics[width=8cm, height=4cm]{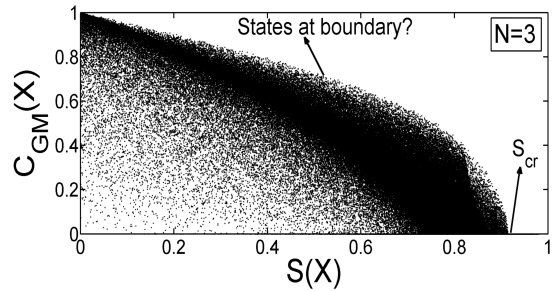}
\caption{(Color online) Genuine multipartite entanglement and entropy plane for three-qubit $X$-states. Each of the million points corresponds to the $C_{GM}(X)$ and $S(X)$ calculated for a randomly generated $X$-density matrix. We are interested in the class of states that lies at the boundary. We also want to know what the critical entropy, $S_{cr}$, is as a function of $N$. Critical entropy is the entropy above which no possible system state in the $X$ form possesses genuine multipartite entanglement.}\label{fig.3qubit}
\end{figure}

In this section, we explore the entropy verses genuine multipartite entanglement properties of $N$-qubit $X$-states. For concreteness, we begin by considering a three-qubit system (The two-qubit system was studied in \cite{Munro-01}). We randomly generate a million $X$-density matrices of three qubits and determine their linear entropy and genuine multi-partite concurrence using Eqs. (\ref{e.S}) and (\ref{e.C_gm}) respectively. The entanglement-entropy plane for the three-qubit $X$-states is shown in Fig. \ref{fig.3qubit} where we have plotted $C_{GM}$ and $S$ for each of the randomly generated $X$ matrices. 
We notice two qualitative trends from Fig. \ref{fig.3qubit}. First is that as the entropy increases, the maximum amount of achievable genuine multi-partite entanglement decreases. The second important thing to notice is that beyond a critical amount of entropy (close to 0.9 in this case), no state possesses genuine multipartite entanglement. These qualitative features can be quantitatively captured by knowing the class of states that lies at the boundary of the entanglement-entropy plane. These are the states that have the maximum entropy for a given amount of entanglement or said in a different way, has the maximum amount of entanglement for a given amount of entropy. We will refer to this class of states as \textit{maximally entangled mixed $X$-states} (X-MEMS). The purpose of this report is to find X-MEMS for an $N$-qubit system. 

Without loss of generality, we assume that for a genuinely multi-partite entangled state the concurrence is given by, $C_{GM}(X)=2(|z_1|-\sum_{j\neq 1}^n\sqrt{a_jb_j})$. We now prove that for every density matrix of the form given in Eq. (\ref{e.X}), we can find another density matrix, $\bar{X}$, that has the same $C_{GM}$ as the density matrix $X$ but has a higher entropy. The density matrix elements of $\bar{X}$ in terms of the matrix elements of $X$ are as follows:
\begin{align}\label{e.matel_barX}
\bar{a}_1&=a_1,\nonumber\\
\bar{a}_i&=a_i+b_i,\quad\quad 2\leq i\leq n,\nonumber\\
\bar{b}_1&=b_1,\nonumber\\
\bar{b}_i,\bar{z}_i&=0,\qquad\quad\quad 2\leq i\leq n,\nonumber\\
\bar{z}_1&=|z_1|-\sum_{j\neq 1}^n\sqrt{a_jb_j}.
\end{align}
The $b_i$'s are added to the $a_i$'s to ensure that the trace of $\bar{X}$ is one. It is clear that the genuine multi-partite entanglement of $X$ and $\bar{X}$ is the same: 
\begin{align}
C_{GM}(\bar{X})=C_{GM}(X). 
\end{align}
In order to prove that the entropy of $\bar{X}$ is more than the entropy of $X$, we note that for $|z_1|\geq\sum_{i}^n\sqrt{a_ib_i}$, we have 
\begin{align}
S(\bar{X})-S(X)&=\left(\frac{2^{N+1}}{2^N-1}\right)\sum_{i\neq 1}^n \Big(|z_{i}|^2+2\sqrt{a_ib_i}(|z_{1}|-\sqrt{a_ib_i}-\sum_{j\neq 1,i}^n\sqrt{a_jb_j})\Big),\nonumber\\
&\geq 0.
\end{align}
The facts that $C_{GM}(\bar{X})=C_{GM}(X)$ and $S(\bar{X})\geq S(X)$ imply that the maximally entangled mixed $X$ states have to be of the $\bar{X}$ form for which all the lower diagonal elements except $b_1$ are zero. 

\begin{figure}
\includegraphics[width=8cm, height=4cm]{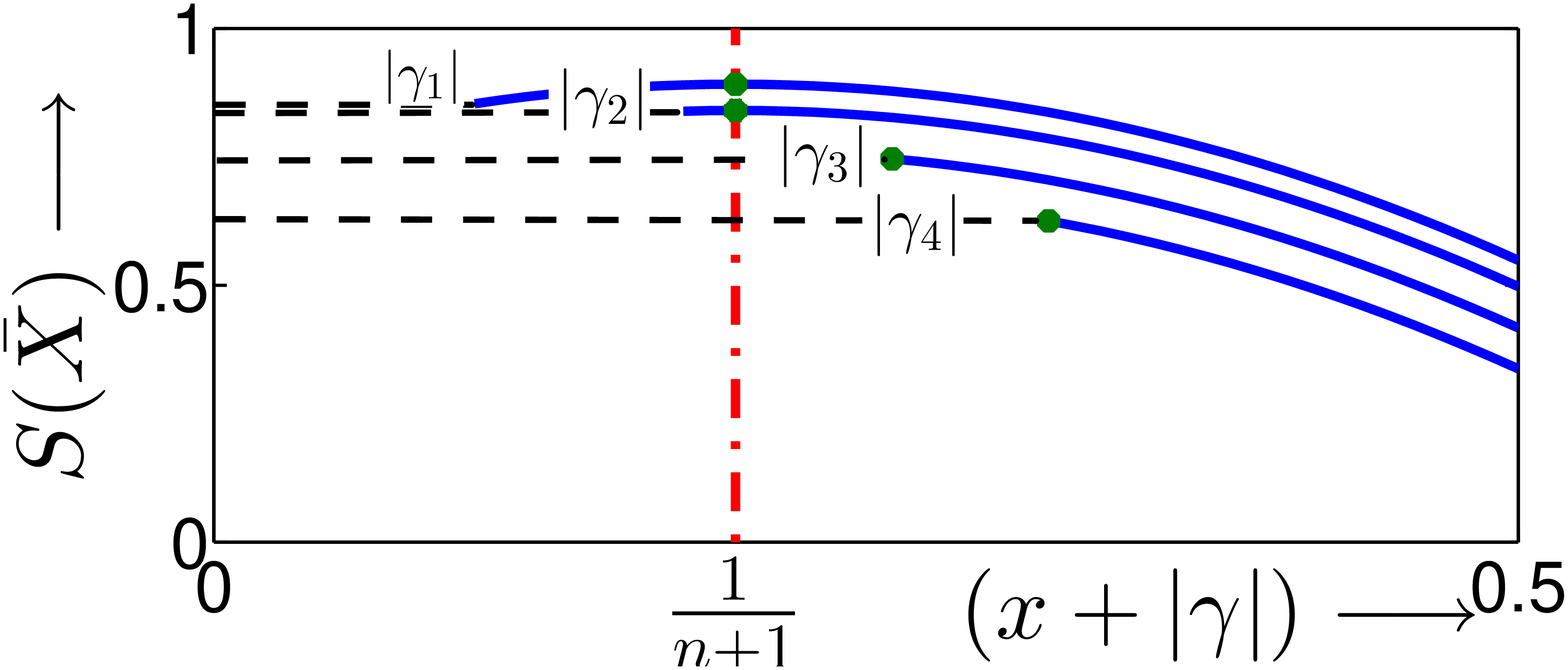}
\caption{(Color online) Maximizing the entropy, $S(\bar{X})$. The solid lines correspond to $S(\bar{X})$ for different values of $|\gamma|$ over the allowed range of $(x+|\gamma|)$. According to Eqs. (\ref{e.constraints}) and (\ref{e.ax}), the allowed range of $(x+|\gamma|)$ is $|\gamma|\leq (x+|\gamma|)\leq 1/2$. The solid dots correspond to the respective maximas of the various curves. We see that the maximum of $S(\bar{X})$ occurs at $(x+|\gamma|)=1/(n+1)$ if $|\gamma|\leq 1/(n+1)$ and at $(x+|\gamma|)=|\gamma|$ if $|\gamma|\geq 1/(n+1)$.}\label{fig.max_S}
\end{figure}

\begin{figure}
\includegraphics[width=8cm, height=4cm]{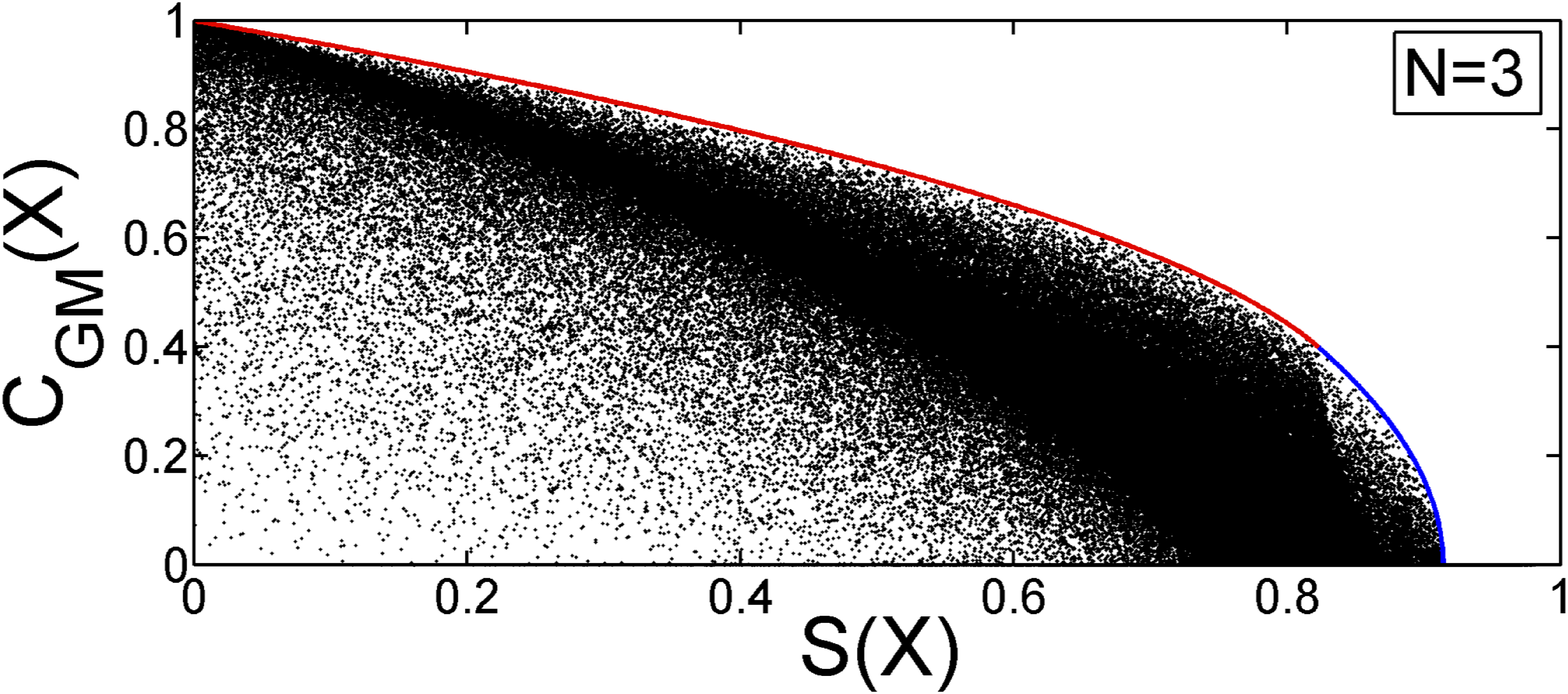}
\includegraphics[width=8cm, height=4cm]{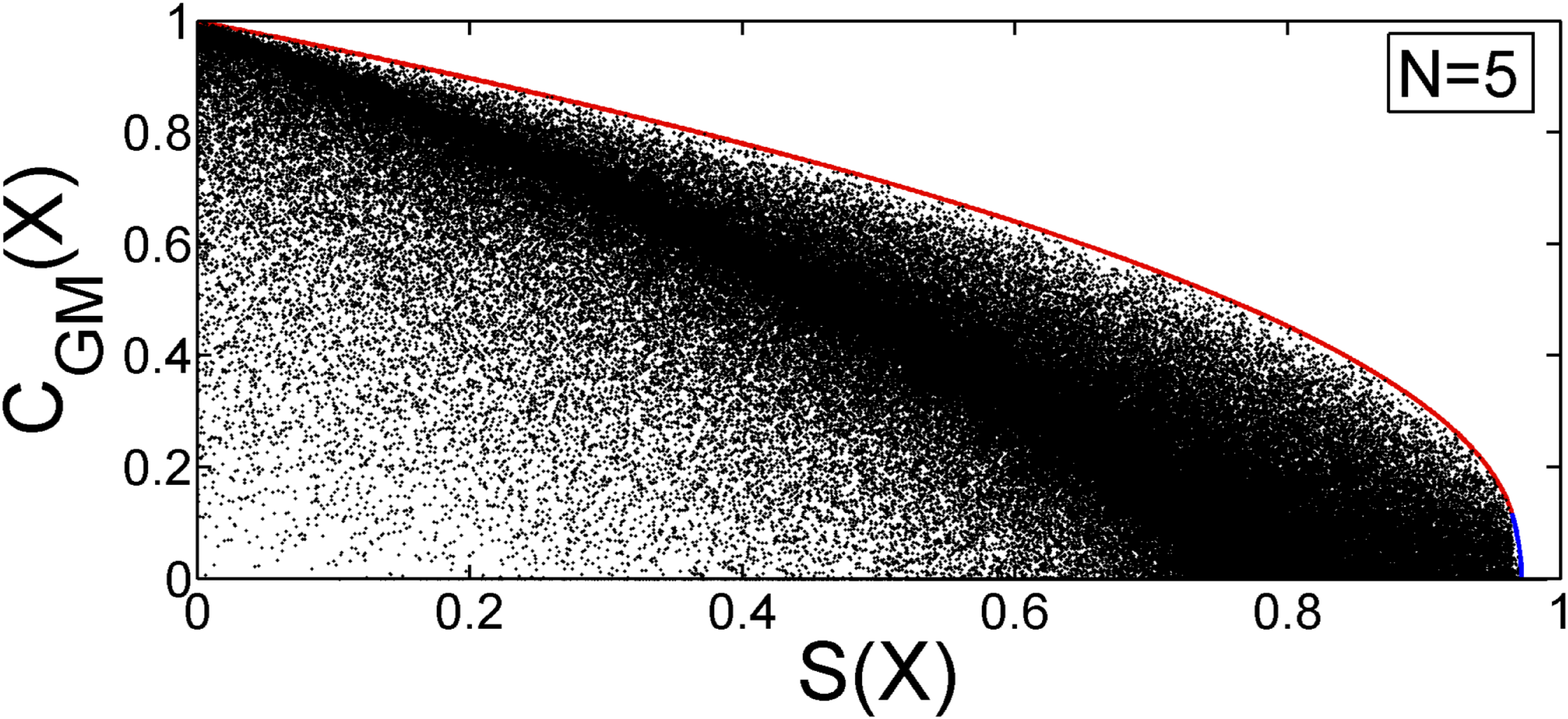}
\caption{(Color online) Entanglement and entropy of randomly generated three-qubit and five-qubit $X$-density matrices. The solid lines correspond to Eq. (\ref{e.Sf}) which analytically gives the maximum entropy for a given amount concurrence. We see that in agreement with the analytical results derived in this report, all the random points lie inside the solid line. Note that as the number of qubits increases from $N=3$ to $N=5$, the critical amount of entanglement beyond which no state is entangled also increases.}\label{fig.5qubit}
\end{figure}

To reduce mathematical complications while deriving the form of X-MEMS, we redefine the following matrix elements of $\bar{X}$ which were previously defined in Eq. (\ref{e.matel_barX}):
\begin{align}
\bar{a}_1&=\alpha+|\gamma|,\nonumber\\
\bar{b}_1&=\beta+|\gamma|,\nonumber\\
\bar{z}_1&=\gamma.
\end{align}
To ensure the positivity and unit trace properties of the density matrix $\bar{X}$, the following conditions must be satisfied:
\begin{align}\label{e.constraints}
\alpha+|\gamma|&\geq0,\nonumber\\
\beta+|\gamma|&\geq0,\nonumber\\
\alpha\beta+(\alpha+\beta)|\gamma|&\geq0,\nonumber\\
\alpha+\beta+2|\gamma|+\sum_{i=2}^n \bar{a}_i&=1.
\end{align}
With this redefinition, the density matrix, $\bar{X}$, takes the following form:
\begin{align}
\bar{X}=
\begin{pmatrix}
\alpha+|\gamma|&&&&&&&\gamma\\
&\bar{a}_2&&&&&0&\\
&&\ddots&&&\reflectbox{$\ddots$}&&\\
&&&\bar{a}_n&0&&&\\
&&&0&0&&&\\
&&\reflectbox{$\ddots$}&&&\ddots&&\\
&0&&&&&0&\\
\gamma^*&&&&&&&\beta+|\gamma|\\
\end{pmatrix},
\end{align}  
For the above form of the density matrix, the genuine multipartite concurrence is simply given by $C_{GM}(\bar{X})=2|\gamma|$. Keeping $|\gamma|$ fixed, we now need to figure out the expressions for $\bar{a}_i$, $\alpha$ and $\beta$ that would maximize the entropy, $S(\bar{X})$. To this end, we note that because of the presence of the off-diagonal element $\gamma$, the diagonal matrix elements $\alpha$ and $\beta$ have to be treated separately than all other diagonal entries, $\bar{a}_i$ for $2\leq i\leq n$. We also note that keeping the off-diagonal element constant, entropy or mixedness is maximized if all the possible diagonal elements are equally populated, i.e. when 
\begin{align}\label{e.ax}
\alpha&=\beta=x,\nonumber\\
\bar{a}_2&=\bar{a}_3=\cdots=\bar{a}_n=y,
\end{align}
where we have defined two new variables, $x$ and $y$. Using the constraint that the sum of the diagonal terms should sum to one, the variable $y$ can be eliminated from the expression of the linear entropy of the density matrix $\bar{X}$ to get:
\begin{align}
S(\bar{X})&= \frac{2^N}{2^N-1}\left(A\left(x+|\gamma|\right)^2+B\left(x+|\gamma|\right)+C\right),
\end{align}
where 
\begin{align}
A&=-2\left(\frac{n+1}{n-1}\right),\nonumber\\
B&=\frac{4}{n-1},\nonumber\\
C&=\left(1-\frac{1}{n-1}-2|\gamma|^2\right).
\end{align}
We now need to maximize the linear entropy, $S(\bar{X})$, for a given amount of $C_{GM}(\bar{X})=2|\gamma|$. While doing so, we have to remember the constraints in Eq. (\ref{e.constraints}). According to Eq. (\ref{e.constraints}), one must have $x,y>0$ and $2(x+|\gamma|)+(n-1)y=1$, which implies that $(x+|\gamma|)\leq1/2$. Keeping these constraints in mind, we plot $S(\bar{X})$ in Fig. \ref{fig.max_S} as a function of $x+|\gamma|$ for four different values of $|\gamma|$. 
We note that when $|\gamma|\leq1/(n+1)$ the maximum of $S(\bar{X})$ occurs at $(x+|\gamma|)=1/(n+1)$, which according to the trace condition means that $y=1/(n+1)$. On the other hand, when $|\gamma|\geq1/(n+1)$, the maximum of $S(\bar{X})$ occurs at $(x+|\gamma|)=|\gamma|$, which in turn implies that $y=(1-2|\gamma|)/(n-1)$. With these considerations, we finally get the following form of the $X$-density matrices that has the maximum amount of entropy for a given amount of genuine multipartite entanglement:
\begin{align}\label{e.MEMXS}
\tilde{X}=
\begin{pmatrix}
f(\gamma)&&&&&&&\gamma\\
&g(\gamma)&&&&&0&\\
&&\ddots&&&\reflectbox{$\ddots$}&&\\
&&&g(\gamma)&0&&&\\
&&&0&0&&&\\
&&\reflectbox{$\ddots$}&&&\ddots&&\\
&0&&&&&0&\\
\gamma^*&&&&&&&f(\gamma)\\
\end{pmatrix},
\end{align}  
where 
\begin{align}
f(\gamma) = \left\{ 
  \begin{array}{l l}
    1/(n+1) & \quad \text{$0\leq|\gamma|\leq1/(n+1)$}\\
    |\gamma| & \quad \text{$1/(n+1)\leq|\gamma|\leq1/2$,}\\
  \end{array} \right.
\end{align}
and
\begin{align}
g(\gamma) = \left\{ 
  \begin{array}{l l}
    1/(n+1) & \quad \text{$0\leq|\gamma|\leq1/(n+1)$}\\
    (1-2|\gamma|)/(n-1) & \quad \text{$1/(n+1)\leq|\gamma|\leq1/2$.}\\
  \end{array} \right.
\end{align}
As noted earlier, these are also the states that have the maximum amount of entanglement for a given amount of entropy. The degree of genuine multipartite entanglement corresponding to these states is $C_{GM}(\tilde{X})=2|\gamma|$ and the corresponding linear entropy is:
\begin{align}\label{e.Sf}
S(\tilde{X})&= \frac{2^N}{2^N-1}\left(A f^2(\gamma)+B f(\gamma)+C\right).
\end{align}
The form of the matrix $\tilde{X}$ derived in Eq. (\ref{e.MEMXS}) in terms of the concurrence, $C_{GM}=2|\gamma|$, is the main result of this report. This class of states is the maximally entangled mixed $X$-states (X-MEMS). To illustrate our analytically derived results, we plot in Fig. \ref{fig.5qubit} the entanglement versus entropy of a million randomly generated three-qubit and five-qubit $X$-density matrices. The analytical bound found in Eq. (\ref{e.Sf}) is plotted as a solid line. In clear agreement with the analytic bound imposed in Eq. (\ref{e.Sf}), we see that all the randomly generated points lie inside the solid line.
 
It should be noted that the general results derived here for the $N$-qubit case reduces correctly to the results derived by Munro et al. for $N=2$ \cite{Munro-01}. We reiterate the fact that the analysis in this report was restricted to a special form of $N$-qubit density matrices that have the $X$ form as given in Eq. (\ref{e.X}). On the other hand, the results of \cite{Munro-01} were valid for all physically allowed two-qubit states and not just for $X$-matrices. 

\begin{figure}
\includegraphics[width=6cm, height=4cm]{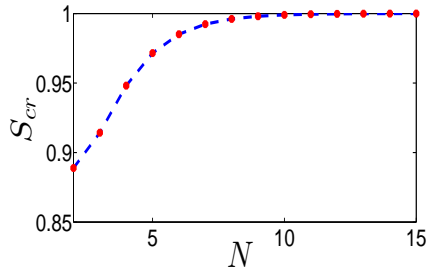}
\caption{(Color online) Critical entropy, $S_{cr}(N)$, as a function of the number of participating qubits, $N$. We see that the critical entropy increases as the number of qubits increases and asymptotically approaches unity.}\label{fig.Scr}
\end{figure}

Putting $|\gamma|=0$ in the Eq. (\ref{e.Sf}), we can get the critical amount of entropy beyond which no $X$-states can exhibit entanglement. This critical entropy as a function of $N$ turns out to be:
\begin{align}\label{e.S_cr}
S_{cr}(N)=\frac{2^{2N-1}}{(2^N-1)(2^{N-1}+1)},
\end{align}
which for $N=2$ correctly gives $S_{cr}(2)=8/9$ \cite{Munro-01}. In Fig. \ref{fig.Scr}, we plot $S_{cr}(N)$ as a function of the number of participating qubits, $N$. 
We see that $S_{cr}(N)$ is an increasing function of $N$ and approaches unity asymptotically. This implies that as the number of qubits increases, we can have states that are more mixed and still exhibit genuine multipartite entanglement.  

\section{Conclusion}\label{s.Conclusion}
In this report, we found the class of $N$-qubit $X$-states that has the maximum amount of genuine multipartite entanglement for a given amount of mixedness. This class of states derived here, referred to as maximally entangled mixed $X$-states (X-MEMS), is an extension to the ones derived for the two-qubit case in \cite{Munro-01}. We find that the critical amount of entropy, $S_{cr}(N)$, above which no $X$-state possesses genuine multipartite entanglement increases as a function of the number of participating qubits, $N$. As the derived class of states has the maximum amount of entanglement for a given amount of entropy, they might have practical utility in cases that require entanglement as a resource but are influenced by unavoidable sources of noise. 

The entire analysis in this report was restricted only to density matrices that have an $X$ form. This restriction was primarily because of the fact that a computable measure which captures genuine multipartite entanglement of an $N$-qubit system exists only for $X$-density matrices. The results in \cite{Munro-01} for the two qubit case were more general. In \cite{Munro-01}, maximally entangled mixed states (MEMS) were found out of all possible two-qubit density matrices and the analysis was not just restricted to density matrices of the $X$ form. The MEMS found in \cite{Munro-01} happens to be of the $X$ form. This fact suggests that the MEMS of an $N$-qubit system might also be of the $X$-form. If this happens to be true, the MEMS of an $N$-qubit system will be the same as the X-MEMS found in this report. Since an analytic form of $C_{GM}$ for an arbitrary $N$-qubit density matrix is yet to emerge, a connection between MEMS and X-MEMS of an $N$-qubit system remains an open question.

For the two-qubit case, it was shown that the maximally entangled mixed states depend upon the measures used to quantify mixedness and entanglement. In this report, the measure of mixedness was chosen to be linear entropy (see Eq. (\ref{e.S})), and the measure chosen to quantify entanglement was genuine multipartite concurrence (see Eq. (\ref{e.C_gm})). Since $C_{GM}$ is the only computable measure available to quantify $N$-qubit entanglement, the dependence of X-MEMS on different measures of entanglement is still an open question.

\begin{acknowledgments}
We are thankful to J.H. Eberly for his encouragement and insightful comments. Financial support was received from NSF PHY-1203931.
\end{acknowledgments}

\end{document}